\documentclass[prb,twocolumn,showpacs]{revtex4}
\usepackage{amsmath}
\usepackage{graphicx}

\begin{document}


\title{DDW Order and its Role in the Phase Diagram of
Extended Hubbard Models}

\author{Chetan Nayak}
\affiliation{Department of Physics and Astronomy, University of California,
Los Angeles, CA 90095-1547}

\author{Eugene Pivovarov}
\affiliation{Department of Physics, California Institute of Technology,
Pasadena, CA 91125}
\date{\today }

\begin{abstract}
We show in a mean-field calculation that phase diagrams remarkably similar
to those recently proposed for the cuprates arise in simple microscopic
models of interacting electrons near half-filling. The models are extended
Hubbard models with nearest neighbor interaction and correlated hopping. The
underdoped region of the phase diagram features $d_{{x^2}-{y^2}}$
density-wave (DDW) order. In a certain regime of temperature and doping, DDW
order coexists with antiferromagnetic (AF) order. For larger doping, it
coexists with $d_{{x^2}-{y^2}}$ superconductivity (DSC). While phase
diagrams of this form are robust, they are not inevitable. For other
reasonable values of the coupling constants, drastically different phase
diagrams are obtained. We comment on implications for the cuprates.
\end{abstract}



\maketitle


\section{Introduction}

The high-$T_c$ cuprates exhibit peculiar behavior when underdoped: the
density-of-states is depleted at low energies, as if some of the degrees of
freedom of the system were developing a gap. This behavior, observed
in optical conductivity,\cite{optics}
NMR,\cite{nmr} angle-resolved photoemission,
\cite{arpes} $c$-axis tunneling,\cite{tunneling}
and specific heat measurements,\cite{spec-heat}
was dubbed the `pseudogap'. The
emergence of the `pseudogap' mimics somewhat the impoverishment of the
low-energy excitation spectrum which accompanies the development of
$d_{{x^2}-{y^2}}$ superconductivity and resembles, more generally,
the type of gap formation which is concomitant
with a large class of order parameters.
However, it does not -- at first glance -- appear to be connected with the
formation of an ordered state. Consequently, it was initially believed that
the pseudogap was a crossover phenomenon, and attempts to describe it
depended on various approximate methods of treating states with local,
fluctuating order.\cite{phase-fluct,gauge}

However, it has recently been proposed that the `pseudogap' state \emph{is\/}
actually a broken-symmetry ordered state, and that the signatures of the
order are subtle enough that the state was able to appear incognito.\cite
{Chakravarty01,Varma99,Kivelson98} In Ref.~\onlinecite{Chakravarty01}, the
$d_{{x^2}-{y^2}}$ density-wave (DDW) state \cite{DDW}
was advanced as a candidate order.
The realization that this is a realistic possibility has led to a
re-examination of the experimental circumstances. Recent \emph{elastic\/}
neutron scattering experiments, which directly probe the symmetries broken
by DDW order -- time-reversal and translation by one lattice spacing --
appear to have observed it.\cite{Mook} A number of other experiments
are consistent with the proposal,\cite{Chakravarty01}
especially measurements of the superfluid density as
a function of doping.\cite{Uemura89}

The experimental situation seems promising, which is strong incentive to
reconsider the theoretical state of affairs. If the `pseudogap' state is,
indeed, an ordered state, then we should be able to study it within
mean-field theory, as we would study the antiferromagnetic state,
superconducting state, or any other ordered state. Mean-field theory is
unlikely to explain the detailed shape of the phase boundary, but one can
hope that it will capture the broad features of the phase diagram, such as
its topology and the basic temperature scales. Deep within any phase, with
$T\rightarrow 0$ and far from any quantum phase transitions, the mean-field
Hamiltonian should be the correct Hamiltonian, although the parameters in it
may need to be renormalized from their mean-field values. Thus it seems
natural to simultaneously study the antiferromagnetic (AF), $d_{{x^2}-{y^2}}$
-wave superconducting (DSC), and DDW order parameters in mean-field theory.
The interplay and possible coexistence of these orders should be
qualitatively and semi-quantitatively explained by mean-field theory. Phase
transitions, quantum or thermal, may not be accurately described in their
asymptotic limits, but the AF, DDW, and DSC phases will, as will possible
phases with coexisting AF, DDW, and DSC orders.

However, there is an immediate problem faced by such a program. What
microscopic Hamiltonian should be used? In the early days of high-$T_c$, it
was hoped that the important physics of strong local repulsion and
superexchange, which is present in the simplest models, such as the Hubbard
and $t-J$ models, would be sufficient to explain all of the interesting
physics of the cuprates. This appears not to be the case. Monte Carlo
studies have not found superconductivity in the Hubbard model,
\cite{MonteCarlo} while Monte Carlo calculations, exact
diagonalization, and density-matrix renormalization-group (DMRG)
calculations give conflicting results for the $t-J$ model.\cite{t-J}
DMRG studies have found that the behavior of $n$-leg
ladders depends sensitively on the strength of, for instance,
second-neighbor hopping,\cite{White99} as have Monte Carlo
studies.\cite{Poilblanc00} Indeed,
some numerical results are sensitively dependent on
boundary conditions,\cite{Hellberg99} which is further indication
of the instability of many of these models to relatively
small changes in the parameters. Furthermore,
the physics of charge-ordering is probably not correctly
described by the $t-J$ model without near-neighbor (and
possibly long-range) Coulomb repulsion.\cite{Kivelson96,Zaanen99,Arrigoni00}
Indeed, it is also clear from experiments that relatively small changes --
such as those associated with substituting Nd for La,\cite{Tranquada95}
which is off the radar screen of the $t-J$ and Hubbard models -- can
radically change at least some aspects of the behavior of these materials.
In short, the detailed form of the underlying Hamiltonian matters.

Fortunately, we are not completely in the dark about the nature of the
microscopic Hamiltonian. Local Coulomb repulsion, both on-site and
near-neighbor, is clearly an important part of the physics. This is known
from microscopic calculations of the Hubbard parameters $t$, $U$, and also
from the fact that the undoped parent compounds are antiferromagnetic
insulators. The other important clue -- which derives entirely from
experiments -- is that the cuprates superconduct. The correct microscopic
model (or models) must support $d$-wave superconductivity when doped away
from half-filling. If the Hubbard and $t-J$ do not -- and it appears that
they do not for $t/U$ small -- then they cannot describe the cuprates fully.

Our strategy will be to take a generalization of the Hubbard model which
includes next-neighbor repulsion and, most importantly, pair-hopping (or
correlated hopping). The pair-hopping term favors superconductivity. Even
when it is relatively small, it stabilizes superconductivity in the Hubbard
model, as we will see. There are a variety of ways in which such a term --
or another term with similar effect -- could arise, either from quantum
chemistry \cite{Appel93, Hirsch89} or in the passage to an effective description such
as the $t-J$ model; in both cases, it is essentially a result of strong local
Coulomb repulsion, as superexchange is. In any event, it appears that such
physics is necessary to stabilize superconductivity, so we will incorporate
it in our model. We will find that such a term also leads to DDW order.

To summarize, we consider a model which is chosen so that it incorporates
the basic physics of strong local repulsion \emph{and\/} so that will have a
phase diagram which includes AF at half-filling and DSC at some finite
doping. We find that it naturally supports DDW order. In mean-field theory,
we find a phase diagram in the temperature-doping plane which resembles the
experimental phase diagram of the cuprates, with the DDW phase boundary
playing the role of the experimental pseudogap onset line. This DDW line
continues into the DSC state, so that the underdoped superconducting state
is characterized by both DSC and DDW orders. At low doping, there is also a
region of coexistence between AF and DDW orders. We comment on the
interpretation of experiments vis-\`{a}-vis our findings.


\section{Model Hamiltonian}

We consider the following bilayer lattice model of interacting electrons:
\cite{Nayak00a}
\begin{equation}
\mathcal{H}=\mathcal{H}_{\text{kin}}\mathcal{+H}_{\text{int}},
\label{eqn:lattice_model}
\end{equation}
where
\begin{multline}
\mathcal{H}_{\text{kin}}=-t_{ij}\sum_{\left\langle i,j\right\rangle }\left(
c_{i\sigma }^{(\lambda )\dagger }c_{j\sigma }^{(\lambda )}+\text{h.c.}\right)
\\
+\frac{t_{\perp }}{16}\sum_i\left( c_{i+\hat{x}+\hat{y},\sigma }^{(1)\dagger
}c_{i\sigma }^{(2)}+c_{i+\hat{x}-\hat{y},\sigma }^{(1)\dagger }c_{i\sigma
}^{(2)}\right. \\
\left. -c_{i\sigma }^{(1)\dagger }c_{i\sigma }^{(2)}-c_{i+2\hat{x},\sigma
}^{(1)\dagger }c_{i\sigma }^{(2)}+x\rightarrow y+1\rightarrow 2+\text{h.c.}%
\right) ,  \label{eqn:H_kin}
\end{multline}
and
\begin{multline}
\mathcal{H}_{\text{int}}=U\sum_in_{i\uparrow }^{(\lambda )}n_{j\downarrow
}^{(\lambda )}+V\sum_{\left\langle i,j\right\rangle }n_i^{(\lambda
)}n_j^{(\lambda )} \\
-t_c\sum_{\substack{ \left\langle i,j\right\rangle ,\left\langle i^{\prime
},j\right\rangle \\i\neq i^{\prime} }}c_{i\sigma }^{(\lambda )\dagger
}c_{j\sigma }^{(\lambda )}c_{j\sigma }^{(\lambda )\dagger }c_{i^{\prime
}\sigma }^{(\lambda )}.
\label{eqn:H_int}
\end{multline}
In the formulas above $t_{ij}$ is hopping with $t_{ij}=t$ for nearest
neighbors, $t_{ij}=t^{\prime }$ for next nearest neighbors and $t_{ij}=0$
otherwise. The other parameters are the
tunneling $t_{\perp }$, the on-site repulsion
$U$, the nearest-neighbor repulsion $V$, and next-nearest-neighbor correlated
hopping $t_c$. The indices $i,j$ correspond to a lattice site, $\sigma $ to
the spin, and $\lambda $ to the layer.

The next-nearest-neighbor correlated hopping term is physically kinetic, but
since it is also quartic, we are going to treat it as interaction. It hops
an electron from $i^{\prime}$ to $j$ when $j$ is vacated by an electron
hopping to $i$. These two hops are correlated by virtue of Coulomb interaction
between the electrons. The presence of this term in the cuprates
has been shown in band-structure calculations.\cite{Appel93}
Correlated hopping has been discussed in Refs.~%
\onlinecite{Wheatley88, Hirsch89, Chakravarty93, Assaad97}
as a possible mechanism of superconductivity,
but it has also been found\cite{Nayak00a}
that it favors DDW order as well.

The tunneling term is momentum conserving.\cite{Chakravarty93,
Andersen95} We consider a CuO$_2$ bilayer because
the pseudogap has been best characterized in bilayer
materials such as YBCO and Bi2212.

To derive a mean-field theory, it is convenient to
take the Fourier transform
of Eq.\ (\ref{eqn:lattice_model}) and regroup the terms.
This task would be particularly simple if there were
only one phase at a given set of
parameters. For example, a DDW reduced Hamiltonian
would look like
\begin{equation}
\mathcal{H}_{\text{DDW}}=-g_{\text{DDW}}\int_{k,k^{\prime }}f\left( k\right)
f\left( k^{\prime }\right) c _{k+Q,\sigma }^{(\lambda )\dagger }c _{k\sigma
}^{(\lambda )}c _{k^{\prime }\sigma ^{\prime }}^{(\lambda )\dagger }c
_{k^{\prime }+Q,\sigma ^{\prime }}^{(\lambda )},
\end{equation}
where $f\left( k\right) =\cos k_x-\cos k_y$ (the lattice spacing has been set
to unit) and the DDW mean-field coupling constant is
\begin{subequations}
\begin{equation}
g_{\text{DDW}}=8V+24t_c.
\end{equation}
Similar values of the mean filed coupling constants can be derived for other
phases as well. Thus, for antiferromagnesm, $d$-wave superconductivity and
$(\pi,\pi)$ charge-density wave we derive:
\begin{align}
g_{\text{AF}} &= 2U, \\
g_{\text{DSC}} &= 12t_c-8V, \\
g_{\text{CDW}} &= 16V+24t_c-2U.
\end{align}
\end{subequations}

In fact, the interaction part of the Hamiltonian Eq.~(\ref{eqn:H_int}) can
be further generalized to include the \emph{interlayer\/} Coulomb
interactions:
\begin{equation}
\mathcal{H}_{\text{int}}^{\prime}=U^{\prime}\sum_in_i^{(\lambda)}
n_j^{(\lambda^{\prime})}+V^{\prime}\sum_{\left\langle i,j\right
\rangle }n_i^{(\lambda )}n_j^{(\lambda ^{\prime})},
\end{equation}
where $\lambda \neq \lambda^{\prime}$. Then for the given interlayer
configuration of the order parameters (antisymmetric for AF and DDW and
symmetric for DSC), the mean field coupling constants become:
\begin{subequations}
\begin{align}
g_{\text{DDW}}&= 8V+8V^{\prime}+24t_c, \\
g_{\text{AF}} &= 2U+2U^{\prime}, \\
g_{\text{DSC}} &= 12t_c-8V+8V^{\prime}.
\end{align}
\end{subequations}

For the opposite configuration (symmetric for AF and DDW and antisymmetric
for DSC) the contributions of $U^{\prime}$, $V^{\prime}$ would be
negative, which is the main reason why such configurations have generally
higher energy and are not observed. On the other hand, the fact that five
interaction terms produce only three phases means that we can have the
same phase diagrams (corresponding to a given set of $g_p$'s) for a range
of values of the interaction constants. In the following section we will
assume that $U^{\prime}=V^{\prime}=0$ so that each phase diagram will
correspond to a unique set of ${U,V,t_c}$.

The total Hamiltonian contains the reduced parts corresponding to these
phases as well as the interactions between the order parameters. However,
since we expect $g_{\text{CDW}}$ to be negative so that the corresponding
order parameter is always zero, we will ignore the term corresponding to
this phase. The final form of the reduced Hamiltonian is
\begin{multline}
\mathcal{H}_{\text{red}}=\int_k\epsilon _{k\lambda \lambda ^{\prime }}c
_{k\sigma }^{\left( \lambda \right) \dagger }c _{k\sigma }^{\left( \lambda
^{\prime }\right) } \\
-g_{\text{AF}}\int_{k,k^{\prime }}c _{k+Q,\sigma }^{(\lambda )\dagger }c
_{k\sigma }^{(\lambda )}c _{k^{\prime }\sigma ^{\prime }}^{(\lambda )\dagger
}c _{k^{\prime }+Q,\sigma ^{\prime }}^{(\lambda )} \\
-g_{\text{DDW}}\int_{k,k^{\prime }}f\left( k\right) f\left( k^{\prime
}\right) c _{k+Q,\sigma }^{(\lambda )\dagger }c _{k\sigma }^{(\lambda )}c
_{k^{\prime }\sigma ^{\prime }}^{(\lambda )\dagger }c _{k^{\prime }+Q,\sigma
^{\prime }}^{(\lambda )} \\
-g_{\text{DSC}}\int_{k,k^{\prime }}f\left( k\right) f\left( k^{\prime
}\right) c _{k\uparrow }^{(\lambda )\dagger }c _{-k\downarrow }^{(\lambda
)\dagger }c _{k^{\prime }\uparrow }^{(\lambda )}c _{-k^{\prime }\downarrow
}^{(\lambda )},  \label{eqn:model}
\end{multline}
where $\epsilon _{k11}=\epsilon _{k22}=\epsilon _k=-2t\left( \cos k_x+\cos
k_y\right) -4t^{\prime}\cos k_x \cos k_y$, $\epsilon _{k12}=\epsilon
_{k21}=\epsilon _{k\perp }=\left( t_{\perp }/4\right) f\left( k\right) ^2$.

The standard Hubbard-Stratonovich mean-field-theoretical treatment of Eq.~(%
\ref{eqn:model}) is to assume the presence of a bosonic mean field, defined
as an order parameter, neglect the fluctuations, find the eigenvalues of the
Hamiltonian and finally, integrate out the fermion degrees of freedom to
derive the free energy.

We define the order parameters of DDW, AF and DSC phases as follows:
\begin{subequations}
\begin{align}
\phi _\lambda &=g_{\text{DDW}}\int_kf\left( k\right) c _{k+Q,\sigma
}^{(\lambda )\dagger }c _{k\sigma }^{(\lambda )}, \\
M_\lambda &=g_{\text{AF}}\int_kc _{k+Q,\sigma }^{(\lambda )\dagger }c
_{k\sigma }^{(\lambda )}, \\
\Delta _\lambda &=g_{\text{DSC}}\int_kf\left( k\right) c _{k\uparrow
}^{(\lambda )\dagger }c _{-k\downarrow }^{(\lambda )\dagger }.
\end{align}
\end{subequations}

We assume that $\phi _\lambda$ and $M_\lambda $
are anti-symmetric in the bilayer index.
Then, the free energy of the system is
\begin{multline}
f=\frac{\left| M\right| ^2}{g_{\text{AF}}}+\frac{\left| \phi \right| ^2}{g_{%
\text{DDW}}}+\frac{\left| \Delta \right| ^2}{g_{\text{DSC}}} \\
+\sum_{s_1,s_2,s_3=\pm 1}\int_{\substack{ k_x>0 \\k_y>k_x }}\left[
s_1\epsilon _k+s_2\epsilon _{k\perp }-\mu \right. \\
-2T\ln \left( 2\cosh \left\{ \frac 1{2T}\left[ \left\{ f\left( k\right)
\Delta \right\} ^2+\left( s_1\left\{ \left[ \epsilon _k+s_2\epsilon _{k\perp
}\right] ^2\right. \right. \right. \right. \right. \\
\left. \left. \left. \left. \left. \left. +\left[ f\left( k\right) \phi
+s_3M\right] ^2\right\} ^{1/2}-\mu \right) ^2\right] ^{1/2}\right\} \right)
\right] .  \label{eqn:free_energy}
\end{multline}

As we expand this expression for small values of the order parameters, we
can construct a Landau-Ginzburg theory:
\begin{multline}
f\left( T\right) =f_0\left( T\right) +\sum_pa_p\left| \Phi _p\right|
^2+\sum_pb_p\left| \Phi _p\right| ^4  \label{eqn:landau-ginzburg} \\
+\sum_{p\neq p^{\prime }}c_{pp^{\prime }}\left| \Phi _p\right| ^2\left| \Phi
_{p^{\prime }}\right| ^2,
\end{multline}
where $p$ denotes the
kind of the order parameter (AF, DDW or DSC) and $\Phi _p$ is the order
parameter ($M$, $\phi $ or $\Delta $, respectively). The $a_p$ coefficients
cross zero at the transitions so that $a_p=0$ are the equations that
determine critical temperature $T_c$:
\begin{equation}
a_p=\frac 1{g_p}-\sum_{s_1,s_2,s_3=\pm 1}\int_{\substack{ k_x>0 \\k_y>k_x }%
}K_p\left( k\right) ,
\end{equation}
where
\begin{subequations}
\begin{align}
K_{\text{AF}}\left( k\right)  &= \frac 1{2\left| \bar{\epsilon}_k\right|
}\tanh \left( \frac{\left| s_1\bar{\epsilon}_k-\mu \right| }{2T}\right) , \\
K_{\text{DDW}}\left( k\right)  &= f\left( k\right) ^2K_{\text{AF}}\left(
k\right) , \\
K_{\text{DSC}}\left( k\right)  &= \frac{f\left( k\right) ^2}{2\left| s_1\bar{%
\epsilon}_k-\mu \right| }\tanh \left( \frac{\left| s_1\bar{\epsilon}_k-\mu
\right| }{2T}\right) .
\end{align}
\end{subequations}
Here $\bar{\epsilon}_k=\epsilon _k+s_2\epsilon _{k\perp }$.
The $b_p$ coefficients are positive:
\begin{equation}
b_p=\sum_{s_1,s_2,s_3=\pm 1}\int_{\substack{ k_x>0 \\k_y>k_x }}K_p^{\prime
}\left( k\right) ,
\end{equation}
where
\begin{subequations}
\begin{align}
K_{\text{AF}}^{\prime }\left( k\right)  &= \frac{\eta _1\left( k\right) }{%
8\left| \bar{\epsilon}_k\right| ^3}, \\
K_{\text{DDW}}^{\prime }\left( k\right)  &= f\left( k\right) ^4K_{\text{AF}%
}^{\prime }\left( k\right) , \\
K_{\text{DSC}}^{\prime }\left( k\right)  &= \frac{f\left( k\right) ^4\eta
_2\left( k\right) }{8\left| s_1\bar{\epsilon}_k-\mu \right| ^3},
\end{align}
\end{subequations}
where
\begin{subequations}
\begin{align}
\eta _1\left( k\right)  &= \tanh \left( \frac{\left| s_1\bar{\epsilon}_k-\mu
\right| }{2T}\right) -\frac{\left| \bar{\epsilon}_k\right| /2T}{\cosh \left(
\frac{\left| s_1\bar{\epsilon}_k-\mu \right| }{2T}\right) ^2}, \\
\eta _2\left( k\right)  &= \tanh \left( \frac{\left| s_1\bar{\epsilon}_k-\mu
\right| }{2T}\right) -\frac{\left| s_1\bar{\epsilon}_k-\mu \right| /2T}{%
\cosh \left( \frac{\left| s_1\bar{\epsilon}_k-\mu \right| }{2T}\right) ^2}.
\end{align}
\end{subequations}
Finally, the $c_{pp^{\prime }}$ coefficients are
\begin{equation}
c_{pp^{\prime }}=\sum_{s_1,s_2,s_3=\pm 1}\int_{\substack{ k_x>0 \\k_y>k_x }%
}K_{pp^{\prime }}^{\prime \prime }\left( k\right) ,
\end{equation}
where
\begin{subequations}
\begin{align}
K_{\text{AF,DSC}}^{\prime \prime }\left( k\right)  &= \frac{f\left( k\right)
^2}{4\left| s_1\bar{\epsilon}_k-\mu \right| ^2\left| \bar{\epsilon}_k\right|
}\eta _2\left( k\right) , \\
K_{\text{DDW,DSC}}^{\prime \prime } &= f\left( k\right) ^2K_{\text{AF,DSC}%
}^{\prime \prime }\left( k\right) , \\
K_{\text{AF,DDW}}^{\prime \prime }\left( k\right)  &= \frac{3f\left(
k\right) ^2}{4\left| \bar{\epsilon}_k\right| ^3}\eta _1\left( k\right) .
\end{align}
\end{subequations}
The fact that all $K_{pp^{\prime }}^{\prime \prime } >0$ implies that the
phases compete with each other.


\section{Phase diagram}

The mean field phase diagram can be derived by
minimizing Eq.\ (\ref{eqn:free_energy}) at fixed doping.

\begin{figure}[tbh]
\includegraphics[width=3.25in]{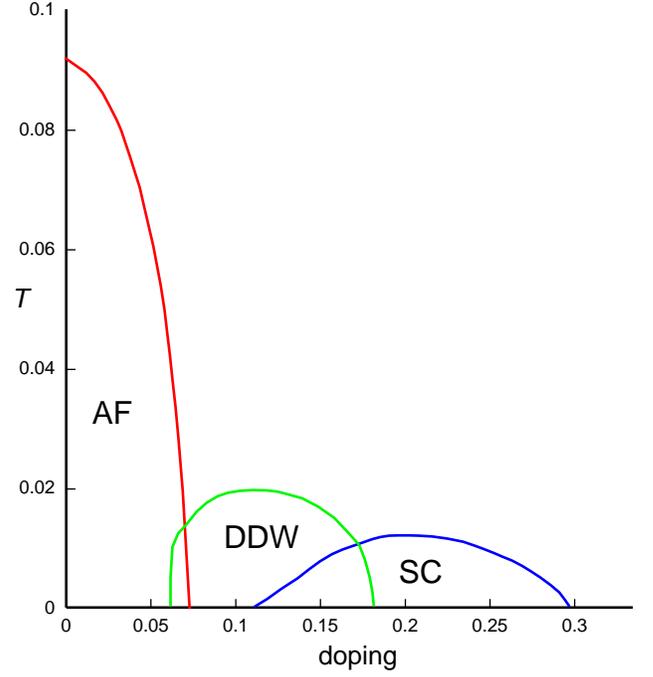}
\caption{Phase diagram for $t=0.5$~eV, $t^{\prime }=-0.025$,
$t_{\perp }=0.05$~eV,
$U\simeq 0.03$~eV, $V=0$~eV, $t_c\simeq 0.8\times 10^{-3}$~eV.
($g_{\text{AF}}=0.06$~eV, $g_{\text{DDW}}=0.02$~eV,
$g_{\text{DSC}}=0.01$~eV.)}
\label{fig:pd1}
\end{figure}
Since there is a large number of parameters in our model,
there is substantial variety in the possible diagrams.
One such diagram, generated with $t=0.5$~eV, $%
t^{\prime }=-0.025$~eV, $t_{\perp }=0.05$~eV, $g_{\text{AF}}=0.06$~eV, $g_{%
\text{DDW}}=0.02$~eV, $g_{\text{DSC}}=0.01$~eV, is shown on the figure \ref
{fig:pd1}. The corresponding values of the interaction constants are $%
U\simeq 0.03$ eV, $V=0$ eV, $t_c\simeq 0.8\times 10^{-3}$ eV. Note that for
these values of the constants, $g_{\text{CDW}}=-0.04$ eV $<0$, which is
consistent with our assumption that a $(\pi,\pi)$
charge-density wave is not energetically favorable.

\begin{figure}[tbh]
\includegraphics[width=3.25in]{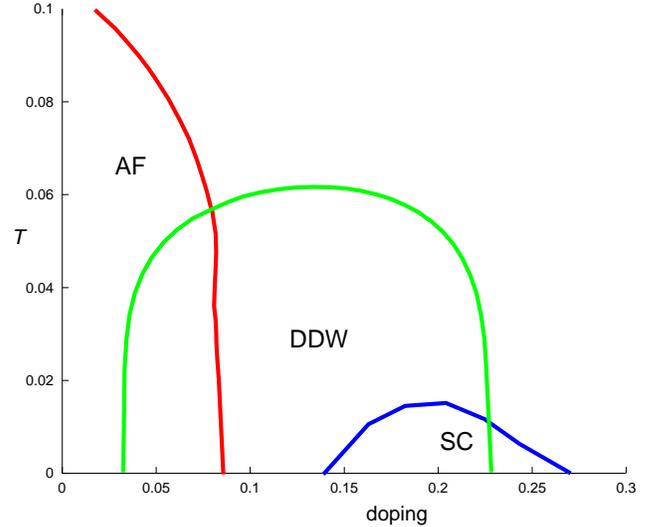}
\caption{Phase diagram for $t=0.5$~eV, $t^{\prime }=0$, $t_{\perp }=0.1$~eV,
$U\simeq 0.042$~eV, $V\simeq 1.7\times 10^{-4}$~eV, $t_c\simeq 1.5\times
10^{-3}$~eV.
($g_{\text{AF}}=0.084$~eV, $g_{\text{DDW}}=0.038$~eV, $g_{\text{DSC}}=0.017$%
~eV.)}
\label{fig:pd2}
\end{figure}
Another diagram, shown on the figure \ref{fig:pd2}, was generated with $t=0.5
$~eV, $t^{\prime }=0$, $t_{\perp }=0.1$~eV, $g_{\text{AF}}=0.084$~eV, $g_{%
\text{DDW}}=0.038$~eV, and $g_{\text{DSC}}=0.017$~eV. The corresponding values
of the interaction constants are $U\simeq 0.042$ eV, $V\simeq 1.7\times
10^{-4}$ eV, $t_c\simeq 1.5\times 10^{-3}$ eV, and also $g_{\text{CDW}%
}\simeq -0.045$ eV.

As we can see, in both diagrams the antiferromagnetic transition temperature
at half-filling is close to 1000 K. This should be understood as the scale at
which two-dimensional antiferromagnetic correlations develop
locally. Due to the Mermin-Wagner-Coleman theorem, which states that
a continuous symmetry cannot be broken spontaneously at finite-temperature
in $2D$, the transition temperature is zero for a single bilayer.
The coupling between different bilayers (which is not
included in our single-bilayer calculation) stabilizes the
antiferromagnetic phase with a transition temperature around 410 K.
In lightly-doped cuprates, the presence of impurities
causes the misalignment of locally ordered antiferromagnetic
clusters, thereby forming a spin glass. Thus, if we interpret
our $T_N$ as the scale of local $2D$ antiferromagnetic order,
which could become $3D$ antiferromagnetic order or
spin glass order, then the phase diagrams of Figs.~\ref{fig:pd1} and
\ref{fig:pd2} are very reasonable, indeed.

\begin{figure}[tbh]
\includegraphics[width=3.25in]{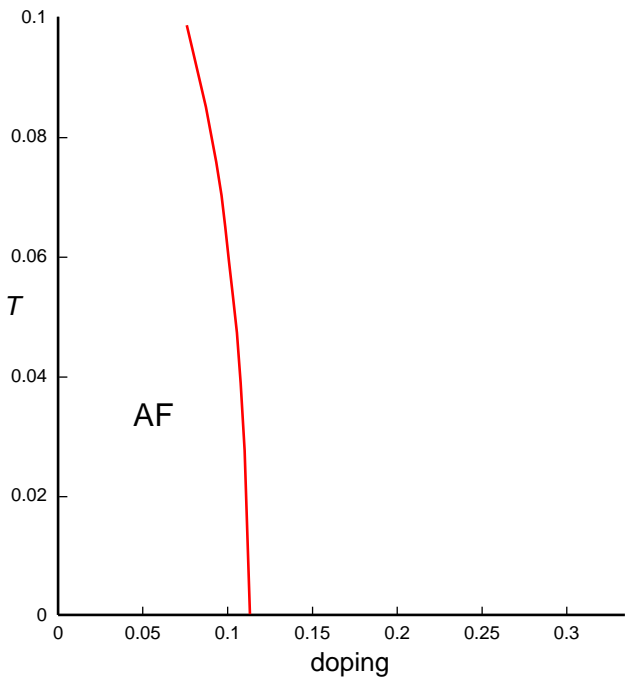}
\caption{Phase diagram for $t=0.5$~eV, $t^{\prime }=-0.025$,
$t_{\perp }=0.05$~eV, $U=0.05$~eV, $V=0$~eV,
$t_c\simeq 3.3\times 10^{-4}$~eV.
($g_{\text{AF}}=0.1$~eV, $g_{\text{DDW}}=0.008$~eV,
$g_{\text{DSC}}=0.004$~eV.)}
\label{fig:pd3}
\end{figure}

Experiments might lead us to expect
that DDW order would occur in the range of doping
between 0.07 and 0.19. This range is smaller than one shown on Fig.~\ref%
{fig:pd2} and a bit larger than that shown in Fig.~\ref{fig:pd1}. The
temperature scale for this phase on Fig.~\ref{fig:pd1} is very reasonable; it is
almost three times higher on Fig.~\ref{fig:pd2}. This change occurred primarily
as a result of the increased value of $t_c$. If we further increase $t_c$ to
$1.9\times 10^{-3}$ eV, the DDW phase will begin to suppress the
AF phase and will expand up to half-filling at finite
temperatures. In general, varying the interaction constants by less that
20~-- 30\% does not change the phase diagram qualitatively. However, larger
variations lead to completely different classes of phase diagrams, such as
those with the AF phase suppressed or
without a DDW phase at all. For example, Fig.~\ref{fig:pd3} shows the case when
due to the smaller value of correlation hopping, both DDW and DSC phases
disappear and only AF phase remains in the diagram.

The DSC phase occupies a doping range away from half-filling
primarily as a result of band structure effects associated
with the bilayer splitting. In the absence of other orders,
it would extend all the way to half-filling, but it is suppressed
at low doping by DDW and AF order. In a more realistic calculation,
superconductivity would be suppressed close to
half-filling by no-double-occupancy constraint,
i.e. by strong local Coulomb repulsion. However, the DSC
phase never even makes it that close to half-filling because
the DDW phase intervenes.

An important feature common to both diagrams is the existence of regions
with two simultaneous kinds of order. Namely, there is a region with DDW+AF
order and a region with DDW+DSC order. The system is an insulator in
the AF state at half-filling, a metal in the DDW and DDW+AF states, and a
superconductor in the DSC and DDW+DSC states.

All of the transitions are of second order at the mean-field level
because of the signs of
the $c_{pp'}$ couplings between the order parameters in
the Ginzburg-Landau theory Eq.~(\ref{eqn:landau-ginzburg}).

The calculated dependence of the chemical potential $\mu$ on the
doping $x$ inside the DDW phase and in its proximity is nonmonotonic.
This is due to the rapid development of the DDW gap,
which causes the chemical potential to be lower than
in the normal state. 
The thermodynamic inequality $(\partial \mu/\partial x)_{T,V}\leq 0$
implies that when this is violated,
mean-field theory should be corrected
using Maxwell's construction, which signals that
fluctuations drive the transitions first-order as a function of
$\mu$. Consequently, we would expect the underdoped side of the DSC
phase to be characterized by a smaller than expected chemical shift,
as has been observed.\cite{Ino97}
A first-order phase transition as a function of chemical potential
is manifested as phase separation in a two-phase coexistence
region spanning a range of dopings when the doping is held fixed instead.
It has been argued that such phase separation will
be precluded by Coulomb interactions, thereby leading to stripe
formation.\cite{Kivelson96,Zaanen99}


\section{Conclusion}

We have studied the phase diagram of a bilayer lattice model using mean
field theory. Since we have focused on ordered phases, this should be a
valid approximation. We found that for certain ranges of the values of the
interaction constants the phase diagram agrees well with the experimentally
observed phase diagram of YBCO if the `pseudogap' is associated with DDW
order. The diagram remains in qualitative agreement with the experimental
data when the parameters of our model vary by less than 20~-- 30\% and becomes
qualitatively different for larger variations. Clearly, such a phase diagram
is reasonably robust, but is hardly inevitable. This is reassuring because
high-temperature superconductivity is stable, but only appears in a special
class of materials (to the best of our current knowledge).

There are some systematic errors associated with mean-field theory, on which
we now comment. It underestimates the effect of fluctuations. Thus, the
N\'eel temperature is very large in mean-field theory, while it should
actually be zero in any strictly two-dimensional system. However, the N\'eel
temperature which we find should be regarded as the temperature below which
a renormalized classical description is valid.\cite{Chakravarty89} The
N\'eel temperature observed in experiments is associated with the crossover
from 2D to 3D. Mean-field theory also overestimates the coupling which
drives antiferromagnetism, which it takes to be essentially $U$. For small $%
U $, this is correct, but for large $U$, it should be replaced by $J\sim{t^2}%
/U $. Indeed, the large-$U$ limit is generally somewhat problematic near
half-filling since the Gutwiller constraint is not enforced in mean-field
theory. The $d_{{x^2}-{y^2}}$ symmetry of the DDW and DSC states lead one to
the erroneous conclusion that they are completely unaffected by large $U$.
This cannot, of course, really be true; clearly, mean-field theory
underestimates the tendency of large-$U$ to push these ordered states away
from half-filling. The seemingly small value of $U$ taken in our calculation
should be interpreted in light of these observations. Other mean-field treatments
which incoporate strong local Coulomb repulsion more prominently
have also found DDW order in a generalization of the $t-J$ model
\cite{Cappelluti99} and in the Hubbard model with nearest neighbor
attraction.\cite{Stanescu01}

We find that the scale associated
with superconductivity is largely determined by the strength of correlated
hopping. At the moment, this is rather \emph{ad hoc}, but we had little
choice but to introduce some term of this sort in order to have a phase
diagram which includes superconductivity. It is possible that the
superexchange coupling $J$ plays a more important role than we have accorded
it in setting $T_c$, but superexchange is beyond a mean-field treatment.

As we have seen, the very term which stabilizes superconductivity also
supports the development of DDW order. One way of interpreting our results
begins with the observation that the DDW order parameter, when combined with
the real and imaginary parts of the DSC order parameter form a triplet under
an $SU(2)$ group of transformations.\cite{Nayak00a,Nayak00b} If this
`pseudospin' $SU(2$) is a symmetry of the Hamiltonian, then DDW and DSC
orders will be equally favored. Thus, one can envision that
the important order-producing term in the Hamiltonian is
$SU(2)$-symmetric while small
symmetry-breaking terms drive the system into either the DDW or DSC states.
Our result shows that pair-hopping is of this form. Are all physically
reasonable mechanisms for $d_{{x^2}-{y^2}}$ superconductivity similarly
invariant under pseudospin $SU(2)$? This is an open problem; we have
answered in the affirmative for one particular class of Hamiltonians.


\begin{acknowledgments}
We would like to thank S.~Chakravarty for discussions.
C.N. is supported by the National Science Foundation
under Grant No. DMR-9983544 and by the Alfred P.
Sloan Foundation. E.P. is supported in part by the Department
of Energy under grant DE-FG03-92-ER\ 40701.
\end{acknowledgments}

\end{document}